\documentclass[graybox]{svmult}

\usepackage{mathptmx}       
\usepackage{helvet}         
\usepackage{courier}        
\usepackage{type1cm}        
\usepackage{makeidx}         
\usepackage{graphicx}        
\usepackage{multicol}        
\usepackage[bottom]{footmisc}
\usepackage{subfigure}
\usepackage{natbib}

\def\Msun{\hbox{M$_{\odot}$}}

\makeindex             

\begin{document}

\title*{An 80 pc Long Massive Molecular Filament in the Galactic Mid-Plane}
\titlerunning{A Massive Molecular Filament}
\author{Cara Battersby \& John Bally}
\institute{CASA, University of Colorado, Boulder, CO 80309 \email{cara.battersby@colorado.edu}}

\maketitle

\abstract{The ubiquity of filaments in star forming regions on a range of scales is clear, yet their role in the star formation process remains in question.  We suggest that there are distinct classes of filaments which are responsible for their observed diversity in star-forming regions.  An example  of a massive molecular filament in the Galactic mid-plane formed at the intersection of UV-driven bubbles which displays a coherent velocity structure ($<$ 4 km s$^{-1}$) over 80 pc is presented.  We classify such sources as Massive Molecular Filaments (MMFs; M $\geq 10^{4}$ \Msun, length $\geq$ 10 pc, $\bigtriangleup$v $\leq$ 5 km s$^{-1}$) and suggest that MMFs are just one of the many different classes of filaments discussed in the literature today.  Many MMFs are aligned with the Galactic Plane and may be akin to the dark dust lanes seen in Grand Design Spirals.
}

\section{Observations}
\label{sec:1}
G32.02$+$0.06, a Massive Molecular Filament (MMF), has a coherent velocity structure over about 80 pc  as traced by $^{13}$CO \citep[The Galactic Ring Survey,][]{jac06}.  The mass of this filament over 80 pc is 2 $\times$ 10$^{5}$ \Msun~in $^{13}$CO and 3 $\times$ 10$^{4}$ \Msun~in 1.1 mm dust continuum emission \citep[The Bolocam Galactic Plane Survey,][]{agu11}.  This MMF exhibits a uniform velocity field ($<$ 4 km s$^{-1}$) over 80 pc and is parallel to the Galactic Plane.  It appears to be a ridge formed at the intersection of two UV-driven bubbles.  

G32.02$+$0.06 has been shaped by older generations of massive stars. At least three 10 to 50 pc diameter bubbles, likely old HII regions, each of which contain at least several massive stars appear to have compressed this cloud and created its various loops and
bends. More details on the large-scale environment of this source are presented in \citep{bat12a}.

\begin{figure}
\subfigure{
\includegraphics[width=0.48\textwidth]{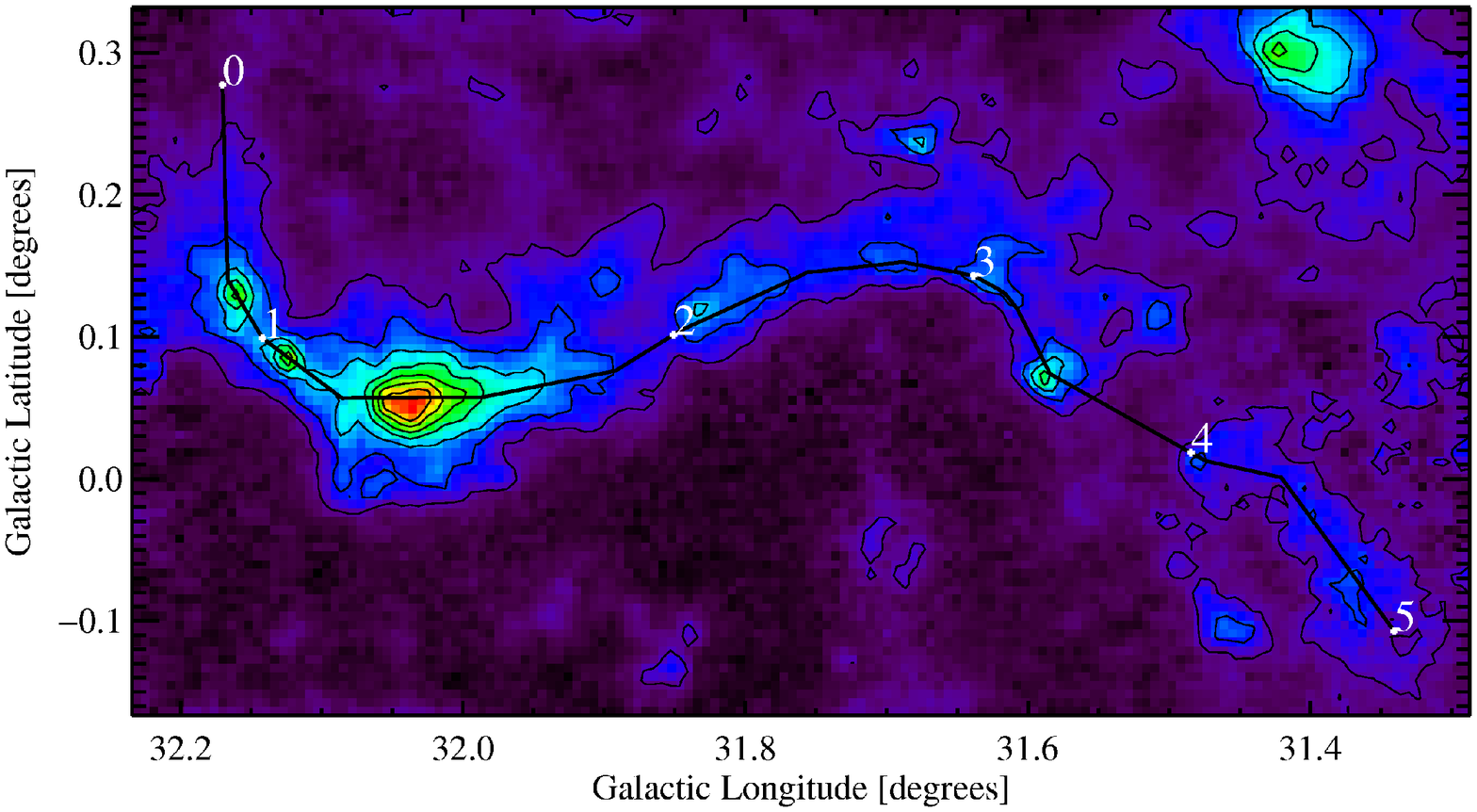}}
\subfigure{
\includegraphics[width=0.5\textwidth]{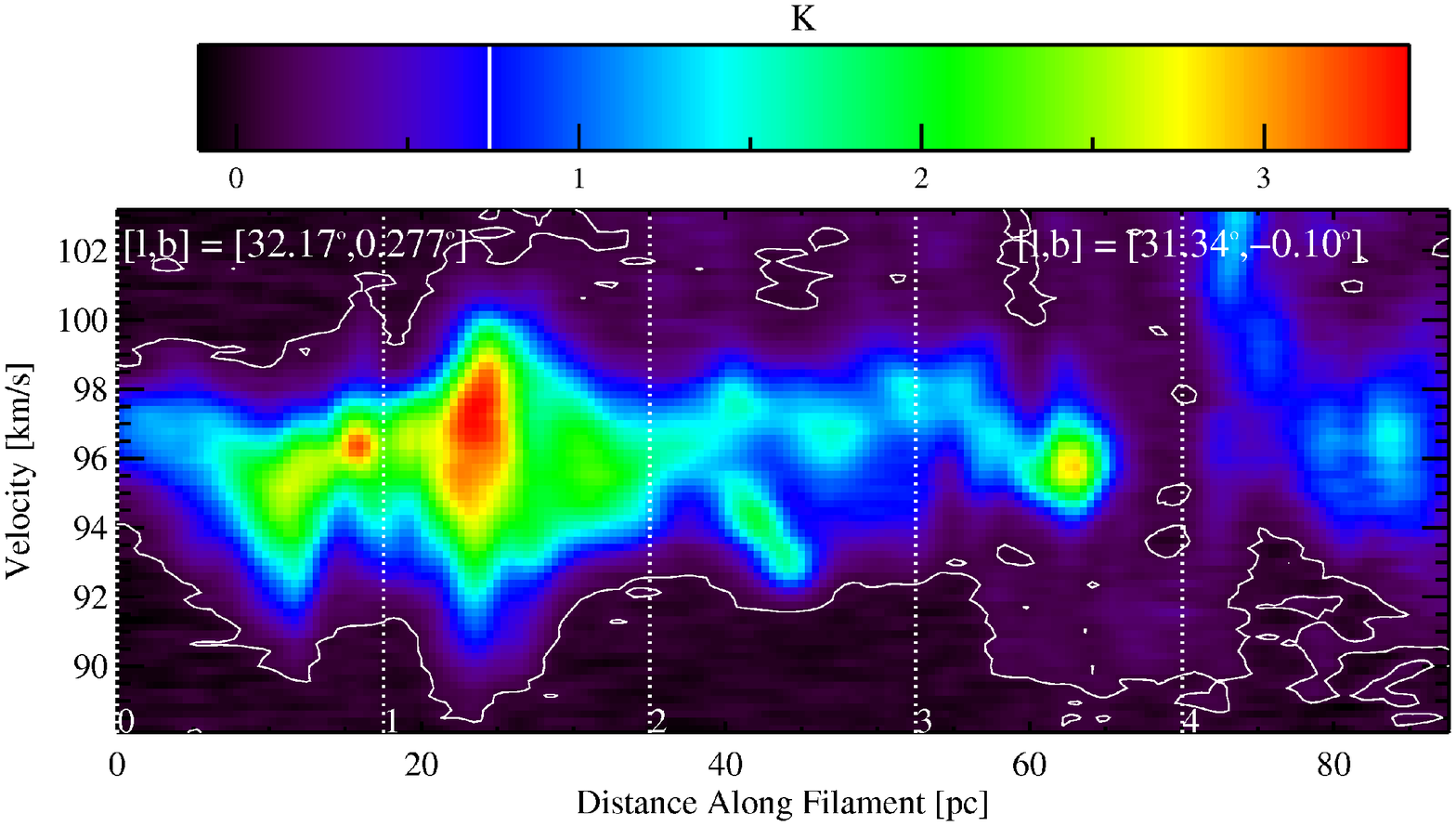}}
\caption{\textit{Left:} An image in $^{13}$CO of the Massive Molecular Filament (MMF) G32.02$+$0.06.  \textit{Right:} A position velocity cut  in $^{13}$CO along the spine of the filament G32.02$+$0.06  as traced by the black line in the left panel.}
\label{fig:1}       
\end{figure}

\section{Interpretation and Discussion}
\label{sec:2}
Nessie \citep{jac10} is another such MMF; it too exhibits coherent ($<$ 3.4 km s$^{-1}$) velocity structure over about 80 pc and has a total mass of about 10$^{4}$ \Msun~as traced by dense gas, comparable to G32.02$+$0.06.  Nessie is also parallel to the Galactic Plane, but slightly offset in latitude from the mid- plane ($|b|$ $\sim$ -0.4$^{o}$).  

There are at least a handful of other MMFs aligned with the Galactic Plane (Tackenberg et al., submitted).  If a majority of MMFs are aligned with the Galactic Plane this indicates that Galactic processes such as shear and spiral density waves are more important than super-bubbles in their formation.  Their alignment with the Galactic Plane is analogous to dark dust lanes along spiral arms seen in face-on galaxies. 

G32.02$+$0.06 represents just one example of an MMF in the Galactic mid-plane formed by the compression of previous generations of massive stars.  We classify such filaments 
(M $\geq 10^{4}$ \Msun, length $\geq$ 10 pc, $\bigtriangleup$v $\leq$ 5 km s$^{-1}$)
as MMFs and suggest that they represent just one category of the oft-discussed ``filaments" in the literature of late.

%

\bibliography{references1}{}
\end{document}